\documentclass[journal]{IEEEtran}

\IEEEoverridecommandlockouts

\usepackage{pifont}
\usepackage[ruled,linesnumbered,vlined]{algorithm2e}
\usepackage{colortbl}
\usepackage{paralist}

\usepackage{cite}
\usepackage{amsmath,amssymb,amsfonts}
\usepackage{graphicx}
\usepackage{textcomp}
\usepackage{xcolor}
\def\BibTeX{{\rm B\kern-.05em{\sc i\kern-.025em b}\kern-.08em
    T\kern-.1667em\lower.7ex\hbox{E}\kern-.125emX}}

\renewcommand{\paragraph}[1]{\vspace{1ex}\noindent
  \textbf{\textit{#1}}\hspace{1ex}}

\newtheorem{definition}{Definition}
\newtheorem{corollary}{Corollary}

\usepackage{url}

\usepackage[scaled=0.88]{helvet}

\usepackage[normalem]{ulem}

\newcommand{\sysname}{CryptoEval}

\begin{document}

\title{\sysname{}: Evaluating the Risk of Cryptographic Misuses in Android Apps with Data-Flow Analysis}


%
%
%
%

\author{Cong~Sun, Xinpeng~Xu, Yafei~Wu, Dongrui~Zeng, Gang~Tan, Siqi~Ma, and~Peicheng~Wang 
\thanks{C. Sun, X. Xu, Y. Wu, and P. Wang are with the School of Cyber Engineering, Xidian University, 710071 China, e-mail: suncong@xidian.edu.cn.}
\thanks{D. Zeng is with Palo Alto Networks, Santa Clara, CA, USA. e-mail: dzeng@paloaltonetworks.com}
\thanks{G. Tan is with the Pennsylvania State University, University Park, PA, USA. e-mail: gtan@psu.edu}
\thanks{S. Ma is with the University of New South Wales, Canberra, Australia. e-mail: siqi.ma@adfa.edu.au}
\thanks{The first two authors contribute equally to this work and are co-first authors.}
\thanks{Corresponding author: Cong~Sun. e-mail: suncong@xidian.edu.cn.}
}

\maketitle

\begin{abstract}
The misunderstanding and incorrect configurations of cryptographic primitives have exposed severe security vulnerabilities to attackers. Due to the pervasiveness and diversity of cryptographic misuses, a comprehensive and accurate understanding of how cryptographic misuses can undermine the security of an Android app is critical to the subsequent mitigation strategies but also challenging. Although various approaches have been proposed to detect cryptographic misuse in Android apps, studies have yet to focus on estimating the security risks of cryptographic misuse. To address this problem, we present an extensible framework for deciding the threat level of cryptographic misuse in Android apps. Firstly, we propose a general and unified specification for representing cryptographic misuses to make our framework extensible and develop adapters to unify the detection results of the state-of-the-art cryptographic misuse detectors, resulting in an adapter-based detection tool chain for a more comprehensive list of cryptographic misuses.
Secondly, we employ a misuse-originating data-flow analysis to connect each cryptographic misuse to a set of data-flow sinks in an app, based on which we propose a quantitative data-flow-driven metric for assessing the overall risk of the app introduced by cryptographic misuses. To make the per-app assessment more useful for app vetting at the app-store level, we apply unsupervised learning to predict and classify the top risky threats to guide more efficient subsequent mitigation. In the experiments on an instantiated implementation of the framework, we evaluate the accuracy of our detection and the effect of data-flow-driven risk assessment of our framework. Our empirical study on over 40,000 apps and the analysis of popular apps reveal important security observations on the real threats of cryptographic misuse in Android apps.
\end{abstract}


\section{Introduction}

Cryptographic algorithms and protocols are the indispensable building blocks for modern distributed computing systems and smartphone apps. Cryptographic primitives, implemented as cryptographic APIs in typical libraries, provide various functionalities such as keys/certificate management, encryption, decryption, digital signature, message digest, and utilization of security protocols. Such cryptographic primitives have been reported vulnerable to many attacks and misused seriously on different platforms \cite{DBLP:conf/apsys/LazarCWZ14, DBLP:conf/ccs/EgeleBFK13, DBLP:conf/nss/LiZLG14, DBLP:conf/codaspy/AlghamdiALM18}.

Previous detection approaches of cryptographic misuses fall into three categories: static analysis \cite{DBLP:conf/ccs/EgeleBFK13, DBLP:conf/ccs/MuslukhovBB18, DBLP:conf/ccs/MaLLD16, DBLP:conf/ecoop/KrugerS0BM18, DBLP:conf/ccs/RahamanXASTFKY19, DBLP:conf/ccs/FahlHMSBF12}, dynamic analysis \cite{DBLP:conf/csfw/FocardiS17}, or a combination of both \cite{DBLP:conf/ndss/SounthirarajSGLK14, DBLP:conf/nss/LiZLG14}. In general, static analysis approaches tend to be more scalable. Unlike the studies that take the common rules of misuse as a threat model \cite{DBLP:conf/ccs/EgeleBFK13, DBLP:conf/ccs/MuslukhovBB18, DBLP:conf/ccs/MaLLD16}, recent studies \cite{DBLP:conf/ecoop/KrugerS0BM18, DBLP:conf/ccs/RahamanXASTFKY19} focused on more fine-grained vulnerability types of cryptographic misuse. To create and maintain the up-to-date rules of misuses from the developers' perspective, the research community has investigated the evolutionary characteristics of cryptographic misuses and their security fixes \cite{DBLP:conf/pldi/PaletovTRV18, DBLP:conf/msr/GaoKLBK19}. Because of the pervasiveness and diversity of cryptographic misuses, one-hop mitigation is impractical. For example, weak hashing may either be used to check the integrity of inconsequential local data or to the password delivered to some network interface. The developer tends to fix the later case as a priority. We have to estimate the real threat of the misuses and decide how we react to the related vulnerabilities, e.g., tolerating, correcting, or applying runtime resistance to them. Currently, the way used to estimate the threat level of cryptographic misuse is simple. The state-of-the-art approaches attach the severity level to the vulnerability type itself. This kind of assessment is coarse-grained and imprecise for measuring the real threat caused by cryptographic misuses. In Android apps, the risk of cryptographic misuse is highly dependent on the data flows affected by the misuse. Suppose a data source operated by some cryptographic misuse flows to a dangerous sink, such as a network sink. In that case, this kind of cryptographic misuses is riskier than those not involved in any sensitive data flow.

This paper proposes an extensible framework for assessing the risk of cryptographic misuses in Android apps. The framework consists of two main components. The first is an adapter-based detection that assembles three state-of-the-art misuse detectors into a detector chain. Each detector in the chain has an adapter to parse the output of the detector, interpret the detected cryptographic misuses into a unified format, and map each misuse to a concrete type in a more comprehensive list of vulnerabilities. The second component is a data-flow-driven risk assessment to quantify the threat level of cryptographic misuses for each app and profile the threat summary to guide the app vetting on a large scale. In this component, we first use a misuse-originating data-flow analysis to obtain the data flows between each misuse and each possible information-leaking channel. Then we quantify the overall security risk for an app based on such data flows. A clustering-based approach is used in our evaluation to predict the most significant threats to apps on a large scale. We highlight our contributions as follows:

\begin{compactitem}[\textbullet]
\item We propose a new detection scheme to improve the precision and recall of static cryptographic misuse detection. The new scheme aims at assembling multiple detectors as a toolchain to detect a more comprehensive list of cryptographic vulnerabilities, which is a more complete threat model in the cryptographic misuse analysis of Android apps.

\item We propose a risk assessment of cryptographic misuses based on misuse-originating data-flow analysis with precise sink category identification. With the identified data flows between cryptographic misuses and the related data leakages, the risk level of misuses is quantified for each app, and the most significant threats of misuses are predicted on a large scale with clustering for the app vetting.

\item With an instantiated implementation of our framework, we evaluate our threat model's effectiveness, the adapter-based detection's accuracy, and the effect of data-flow-driven risk assessment on over 40,000 real-world apps, including several popular apps. Our security observations indicate several important risk patterns.
\end{compactitem}

\section{Motivating Example}\label{sec:example}

\begin{figure}[t]
  \centering
  \includegraphics[width=\linewidth]{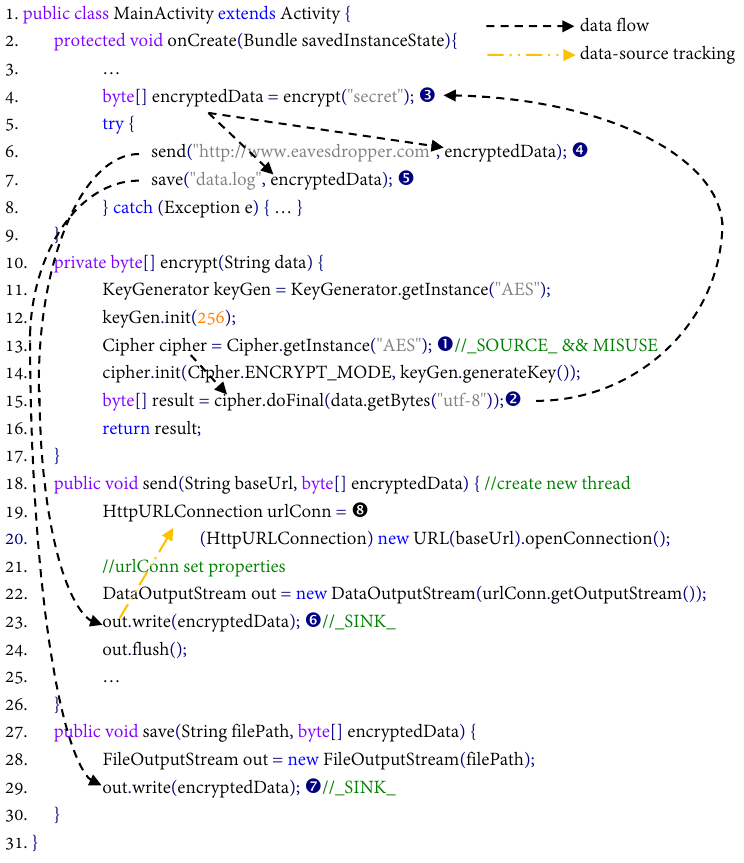}
  \caption{Motivating Example}\label{fig:example}
  \vspace{-3ex}
\end{figure}

The example in Figure~\ref{fig:example} shows that different data flows originating from cryptographic misuse may lead to different impacts and severities to the application. In our work, data flow refers to the data propagation from a source method invocation to a sink method invocation. In method \textsf{encrypt}, we firstly derive the encryption key with \textsf{KeyGenerator}. Then we use \textsf{Cipher.getInstance} to construct and initialize the \textsf{Cipher} object. Finally, we call \textsf{Cipher.doFinal} to encrypt the sensitive string and return the ciphertext. The parameter of \textsf{Cipher.getInstance} should consist of three parts: \textit{algorithm}, \textit{mode of operation}, and \textit{padding scheme}. When we only specify the algorithm ``\textsf{AES}'', the method will apply the ECB mode of operation by default, which has been proved insecure. The state-of-the-art approach~\cite{DBLP:conf/ccs/RahamanXASTFKY19} straightforwardly identifies the risk level of this misuse as \emph{medium} severity. In contrast, we want to further understand how this misuse may be exploited to characterize its severity better. If we treat this misused call to \textsf{Cipher.getInstance} as a data source, there are two sensitive data flows in the program: \ding{182}$\rightarrow$\ding{183}$\rightarrow$\ding{184}$\rightarrow$\ding{185}$\rightarrow$\ding{187} and \ding{182}$\rightarrow$\ding{183}$\rightarrow$\ding{184}$\rightarrow$\ding{186}$\rightarrow$\ding{188}. We want to classify the sinks of these flows into different categories to distinguish the severity of the two sensitive data flows. However, identifying the precise category of some sink might take much work. For example, the category of \textsf{DataOutputStream.write}, i.e., \textsf{NETWORK} in our categorization, is related to the type of \textsf{urlConn}. We apply intra-procedural data-source tracking, i.e., \ding{187}$\rightarrow$\ding{189}, to find such correlation and help categorize the sink. As we realized, the two sensitive flows serve as potential attack surface to exploit the misuse in the program. However, different flow categories have different risk potentials based on public evidence. As we have estimated the CVSS \cite{cvss} of the sink categories, the risk rankings in Table~\ref{tab:sink-weight} indicate that the data flow to \textsf{NETWORK}(\ding{187}) tends to be of higher severity than the device-local data flow to \textsf{FILE}(\ding{188}). We assign risk weights to different categories of sinks according to their severity so that the app's overall risk can be quantified more precisely based on the identified \emph{misuse-originating data flows}.

\begin{figure*}[!ht]
  \centering
  \includegraphics[width=0.9\linewidth]{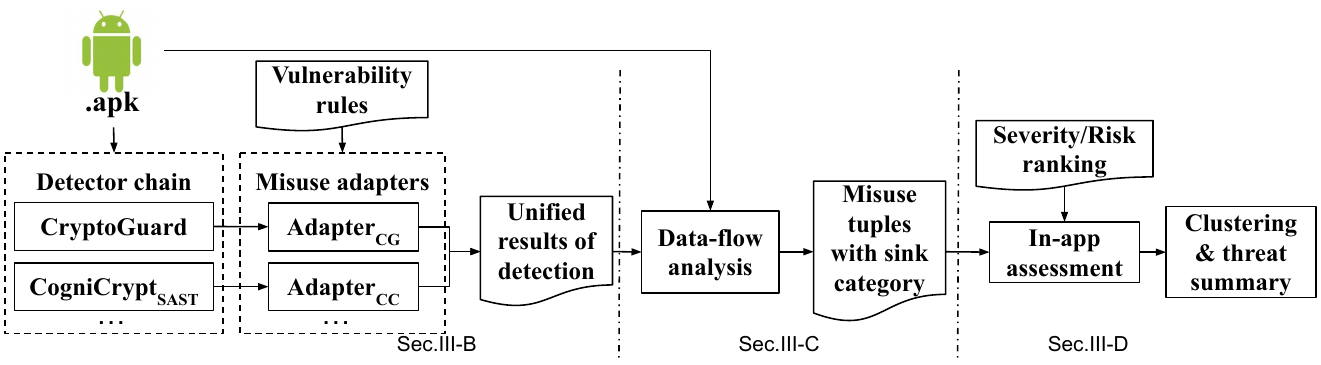}
  \caption{Framework of \sysname{}}\label{fig:framework}
  \vspace{-1ex}
\end{figure*}

\section{\sysname{} Design}\label{sec:approach}

This section describes our framework for assessing the threat level of cryptographic misuses in Android apps.

\subsection{Overview}

We present the framework of \sysname{} in Figure~\ref{fig:framework}. Firstly, our approach assembles the state-of-the-art cryptographic misuse detectors, e.g., \cite{DBLP:conf/ccs/RahamanXASTFKY19, DBLP:conf/ecoop/KrugerS0BM18, DBLP:conf/ccs/MuslukhovBB18}, as a toolchain to detect a more comprehensive list of vulnerability types and to achieve high detection accuracy, which is accomplished by developing a series of adapters. A typical adapter translates the output of a detector into a unified format whose elements will be formally defined in Section~\ref{subsec:adapter}.

\begin{table*}
\renewcommand{\arraystretch}{1.1}
  \caption{Comprehensive list of vulnerabilities caused by cryptographic misuses. Abbreviation of each approach: CG=CryptoGuard\cite{DBLP:conf/ccs/RahamanXASTFKY19}, CL=CryptoLint\cite{DBLP:conf/ccs/EgeleBFK13}, BS=BinSight\cite{DBLP:conf/ccs/MuslukhovBB18}, CR=CDRep\cite{DBLP:conf/ccs/MaLLD16}, CC=CogniCrypt$_\text{SAST}$\cite{cognicrypt}, FD=FixDroid\cite{DBLP:conf/ccs/NguyenWA0WF17}, DC=DiffCode\cite{DBLP:conf/pldi/PaletovTRV18}}
  \label{tab:rules}\footnotesize
  \centering
  \begin{tabular}{rllllll}
    \hline
    ID & Vulnerability Description of \sysname{} & \textsf{CG} & \textsf{CL}/\textsf{BS}/\textsf{CR}$^\dagger$ & \textsf{CC} & Table~1,\textsf{FD} & Fig~9,\textsf{DC} \\
    \hline
    1 & Predictable/constant cryptographic keys & vul.1 & Rule~3 & RequiredPredicateError & Pitfall~3 & R10 \\
    2 & Predictable/constant passwords for PBE & vul.2 & N/A & NeverTypeOfError & N/A & N/A \\
    3 & Predictable/constant passwords for KeyStore & vul.3 & N/A & NeverTypeOfError & N/A & N/A \\
    4 & Custom Hostname verifiers to accept all hosts & vul.4 & N/A & N/A & Pitfall~6 & N/A \\
    5 & Custom TrustManager to trust all certificates & vul.5 & N/A & N/A & Pitfall~11 & N/A \\
    6 & Custom SSLSocketFactory w/o manual Hostname verification & vul.6 & N/A & N/A & N/A & N/A \\
    7 & Occasional use of HTTP & vul.7 & N/A & N/A & Pitfall~8,9 & N/A \\
    8 & Usage of expired protocol by SSLContext & N/A & N/A & ConstraintError & N/A & N/A \\
    9 & Predictable/constant PRNG seeds & vul.8 & Rule~6 & TypestateError & Pitfall~7 & R12 \\
    10 & Cryptographically insecure PRNGs (e.g., java.util.Random) & vul.9 & N/A & RequiredPredicateError & N/A & R3,R6 \\
    11 & Static Salts in PBE & vul.10 & Rule~4 & RequiredPredicateError & N/A & R11 \\
    12 & ECB mode in symmetric ciphers & vul.11 & Rule~1 & ConstraintError & Pitfall~1,5 & R7 \\
    13 & Static initialization vectors (IVs) in CBC mode symmetric ciphers & vul.12 & Rule~2 & RequiredPredicateError & Pitfall~2 & R9 \\
    14 & Fewer than 1,000 iterations for PBE & vul.13 & Rule~5 & ConstraintError & Pitfall~4 & R2 \\
    15 & 64-bit block ciphers (e.g., DES, IDEA, Blowfish, RC4, RC2) & vul.14 & N/A & ConstraintError & N/A & R8 \\
    16 & Insecure asymmetric ciphers (e.g, RSA, ECC) & vul.15 & N/A & ConstraintError & N/A & N/A \\
    17 & Insecure cryptographic hash (e.g., SHA1, MD5, MD4, MD2) & vul.16 & Rule~7$^\dagger$ & ConstraintError & N/A & R1 \\
    18 & Incorrect sequence of cryptographic API calls & N/A & N/A & TypestateError & N/A & N/A \\
    19 & Usage of forbidden APIs & N/A & N/A & ForbiddenMethodError & N/A & R4 \\
    20 & Incomplete usage of cryptographic API & N/A & N/A & IncompleteOperationError & N/A & N/A \\
    21 & Suspected usage needs further testing & N/A & N/A & ImpreciseValueExtractionError & N/A & N/A \\
    \hline
  \end{tabular}
\vspace{-2ex}
\end{table*}

Then, according to the results of cryptographic misuse detection, we perform a misuse-originating data-flow analysis to establish the dependencies between the vulnerabilities caused by the misuses and different sink categories (Section~\ref{subsec:taint-analysis}). This analysis can take advantage of the state-of-the-art context-sensitive and object-sensitive taint analysis on Android apps, i.e., FlowDroid\cite{DBLP:conf/pldi/ArztRFBBKTOM14}, whose sources and sinks are easily configurable. For each app, we update the list of taint sources based on the result of misuse detection. Then, a customized data-flow analysis is conducted from both parameter-related misuses and data-related misuses to find all the specific sinks that may leak data affected by the misuses. Then we use intra-procedural data-source tracking started from these sinks to infer the exact sink categories. We attach the sink categories to each misuse to build explicit correlations for the risk assessments.

We quantify the overall threat level of each app with the detectable vulnerabilities and the data-flow entry points of the vulnerabilities on the app's attack surface. To extend the usage of our per-app assessment, we investigate the potential threat patterns of apps on a large scale for subsequent app vetting at app stores. We extract the features from the misuse tuples with sink information and use unsupervised learning to predict the potential representative threats. An association rule mining is conducted to infer the correlations between different sensitive data flows (Section~\ref{subsec:clustering}).

\subsection{Adapter Construction}\label{subsec:adapter}

Firstly, we investigate the state-of-the-art approaches to analyzing cryptographic misuses \cite{DBLP:conf/ccs/RahamanXASTFKY19, DBLP:conf/ecoop/KrugerS0BM18, DBLP:conf/pldi/PaletovTRV18, DBLP:conf/ccs/NguyenWA0WF17, DBLP:conf/ccs/EgeleBFK13, DBLP:conf/ccs/MaLLD16, DBLP:conf/ccs/MuslukhovBB18}. In these works, the typical features of the vulnerabilities caused by cryptographic misuses, as well as the principles of cryptographic usage are defined. Based on the investigation, we summarized a more comprehensive list of vulnerabilities caused by cryptographic misuses, as shown in Table~\ref{tab:rules}. These vulnerabilities are reported as an extension of the vulnerabilities in \cite{DBLP:conf/ccs/RahamanXASTFKY19}, i.e., the same vulnerability has the same description. The new list of vulnerabilities covers most of the cryptographic misuses claimed by these works. We find that CryptoGuard and \textsc{CogniCrypt}$_\textsc{sast}$ address more vulnerability types in this list than the other approaches. Meanwhile, several approaches have addressed vulnerabilities caused by other reasons. For example, FixDroid \cite{DBLP:conf/ccs/NguyenWA0WF17} also mentioned the security coding pitfalls of SQL injection and local HTML loading. Some rules elicited by DiffCode \cite{DBLP:conf/pldi/PaletovTRV18}, e.g., using BouncyCastle library (R5) and key integrity check (R13), are too restrictive to be a common type of cryptographic misuses; therefore, we do not mention them in our vulnerability list.

According to the ability of different detection approaches presented in Table~\ref{tab:rules}, integrating existing detection tools as a detector chain is sufficient to detect all the vulnerability types in our list. The extensibility of this detector chain is decided by the misuse types covered by the detection approaches, as well as the informativeness of the output of detectors. To represent the cryptographic misuse in a unified form, we formally define the sink-related cryptographic misuse as follows.

\begin{definition}[sink-related cryptographic misuse]
A \emph{sink-related cryptographic misuse} is a tuple $\langle m,id,p,d,t,\textsf{S}\rangle$. It consists of
\begin{compactitem}[\textbullet]
\item $m$: the signature of misused cryptographic API. Generally, the definition of this method is provided by the cryptographic libraries.
\item $id$: the $id$ of vulnerability caused by the misuse, as defined in Table~\ref{tab:rules}.
\item $p$: the signature of the parent method, i.e., the user-defined method where the misuse is located in.
\item $d$: the descriptions of misuse, including the information about the reason and actual parameters of the misuse.
\item $t$: the tool used to detect the misuse, e.g., \textsf{CG}, \textsf{CC}, \textsf{BS}, etc.
\item \textsf{S}: the list of sink categories. Each of the sink categories contains at least one sink that is detected to release sensitive data affected by the misused cryptographic API call.
\end{compactitem}
\end{definition}

Each adapter we construct is a batch-job tool that takes the output of a cryptographic misuse detector as input and produces a set of these tuples. The tuples parsed from the output of detectors hold empty \textsf{S} until the sink categories are detected in the data-flow analysis in Section~\ref{subsec:taint-analysis}.

\begin{corollary}[valid detector chain]\label{corollary1}
A set of cryptographic misuse detectors is \emph{valid} to our approach, if and only if the following conditions are satisfied:
\begin{compactitem}[\textbullet]
\item the union of vulnerability types addressed by each detector contains our comprehensive list of vulnerability types.
\item the output of each detector can be parsed to a set of misuse tuples.
\end{compactitem}
\end{corollary}

We use CryptoGuard and \textsc{CogniCrypt}$_\textsc{sast}$ as the minimal valid detector chain and discuss the use of an extension of the detector chain in RQ1 of Section~\ref{sec:evaluation}. Deciding the element $id$ of each tuple involves mapping from the misuse types of existing approaches to the vulnerabilities in our list. For the detection output of some approaches, e.g., CryptoGuard and BinSight, such mapping is an easy one-to-one case; see \textsf{CG} and \textsf{BS} in Table~\ref{tab:rules}.
However, for the output of \textsc{CogniCrypt}$_\textsc{sast}$, the misuse types are more coarse-grained than our vulnerability list. Therefore, mapping the misuses in their types, e.g., the misuse with type constantError$\in Rule_{\textsf{CC}}$, to our vulnerability types needs more work which will be presented in Section~\ref{sec:implementation}.

\subsection{Misuse-originating Data-Flow Analysis}\label{subsec:taint-analysis}

This section introduces a misuse-originating data-flow analysis to find the sensitive data flows originating from the cryptographic misuses and their related sink categories. Firstly, we investigate the functionality of cryptographic APIs and divide them into two categories:
\begin{compactitem}[\textbullet]
\item Data-related APIs (DAPI): the APIs used for direct data processing, e.g., encryption, message digest, MAC, SSL/TLS operation.
\item Parameter-related APIs (PAPI): the APIs used for constructing the parameters required by DAPI, e.g., PRNG, key generation, and certificate management.
\end{compactitem}

\noindent We manually classify the APIs in JCA cryptographic libraries through a careful study of the JCA API document, and the results are given in Table~\ref{tab:api-classification}. The usage of data-related APIs usually depends on the usage of some parameter-related APIs. For example, in Figure~\ref{fig:example}, if we initialize a 64-bit AES \textsf{KeyGenerator} by the PAPI \textsf{init} on Line 12, which leads to a misuse of weak key, the DAPI \textsf{Cipher.doFinal} on Line 15 and its encrypted data are affected to be insecure. Our data-flow analysis should also address such dependency and take the DAPI as misuse. Specifically, the sources used in our misuse-originating data-flow analysis are firstly expanded based on the relation of PAPIs and DAPIs in the specific app.

\begin{definition}[taint connection analysis]
Given the predefined taint sources $O$ and sinks $S$, a taint connection analysis $TA_{O,S}$ is a procedure deriving a set of code-location pairs from an app. $TA_{O,S}: X\mapsto P(Loc\times Loc)$ such that for each app $x\in X$, $TA_{O,S}(x)=\{(l_o,l_s)|l_o,l_s$ are respectively call sites of $o\in O$ and $s\in S$, and $l_s$ is tainted by setting $l_o$ as taint source$\}$, where $Loc$ holds all the code locations of $x\in X$.
\end{definition}

\begin{definition}[misuse refinement for source]
For a misused cryptographic API, the refinement procedure $\mathcal{R}$ of misuse derives a list of cryptographic APIs on each app $x\in X$:
\begin{displaymath}
\mathcal{R}(m)=
\begin{cases}
\{m\} ,\hfill \text{iff } m\in \text{DAPI}, \\
\{m\}\cup \{s\mid (\_,l_s)\in TA_{\{m\},\text{DAPI}}(x)\}, \hfill \text{iff }m\in \text{PAPI}.\\
\end{cases}
\end{displaymath}
\end{definition}

\begin{algorithm}[t]
\small
\SetKwInOut{Input}{input}\SetKwInOut{Output}{output}

\Input{1) $\Gamma_x$: The misuse tuples of application $x$ generated by the adapters of detector chain. 2) The sink list $S$.}
\Output{The updated misuse tuples $\Gamma_x$.}
\Begin{
  \ForEach{$(\gamma\equiv\langle m,id,p,d,t,\emph{\textsf{S}}\rangle)\in \Gamma_x$}{
    $O\leftarrow \mathcal{R}(\gamma.m)$\;
    $S_{tmp}\leftarrow \{l_s\mid (l_o,l_s)\in TA_{O,S}(x) \wedge l_o\text{ in }\gamma.p\}$\;
    \tcp{Here $\gamma.\textsf{S} = \emptyset$}
    \ForEach{$l_s\in S_{tmp}$}{
      $\gamma.\textsf{S}\leftarrow \gamma.\textsf{S}\cup DSTrack(l_s)$\;
    }
  }
}
\caption{Misuse-originating Data-Flow Analysis}\label{algo:dfa}
\end{algorithm}

The procedure of our data-flow analysis is presented in Algorithm~\ref{algo:dfa}. When the misused API is a parameter-related API, we further consider the data-related APIs tainted by the misused parameter as taint sources. Note that the taint analysis tool can only start taint tracking from all call sites of a specified API method. However, only some of the call sites are misused. Thus, after the taint analysis, we rely on the parent method $p$ of each misuse tuple to filter out taint sources that are not reported as vulnerabilities. Finally, we get all the related sink locations in $S_{tmp}$. Each sink location should fall into a sink category. To identify the particular category for each sink location, we perform a backward intra-procedural data-source tracking $DSTrack$ initiating from each call site of the affected sink to find all the related data types and calculate the exact sink category. For example, from the call site of \textsf{write} in method \textsf{send} of Figure~\ref{fig:example}, we track the data types \textsf{DataOutputStream}, \textsf{byte[]}, \textsf{HttpURLConnection}, and \textsf{String}. Because we have tagged several related classes with the sink category, e.g., \textsf{HttpURLConnection} with \textsf{NETWORK}, we merge the tracked sink categories with the default sink category of \textsf{write}, i.e., \textsf{OUT\_STREAM}. We take the more sensitive category, i.e., \textsf{NETWORK} in \{\textsf{NETWORK, OUT\_STREAM}\}, as the category of this sink and add this sink category to the misuse tuple.

\begin{table}[t]
\renewcommand{\arraystretch}{1.1}
  \caption{Classification of Cryptographic APIs}\small
  \label{tab:api-classification}
  \centering
  \begin{tabular}{lrr}
    \hline
    Package Name & \# DAPI & \# PAPI\\
    \hline
    \textsf{java.security} & 202 & 806 \\
    \textsf{javax.crypto} & 96 & 179 \\
    \textsf{javax.net.ssl} & 99 & 162 \\
    \textsf{javax.xml.crypto} & 23 & 169 \\
  \hline
\end{tabular}
\vspace{-2ex}
\end{table}

\begin{table}[!t]
\renewcommand{\arraystretch}{1.1}
  \caption{Sink Category and Statistics}\small
  \label{tab:sink-category}
  \centering
  \begin{tabular}{lrrr}
    \hline
    Sink Category ($sc$) & \# FlowDroid(default) & \# Added & \#Total\\
    \hline
    \textsf{FILE} & 3 & 0 & 3 \\
    \textsf{LOG} & 14 & 0 & 14\\
    \textsf{NETWORK} & 17 & 1 & 18 \\
    \textsf{SMS\_MMS} & 3 & 0 & 3 \\
    \textsf{SYNC} & 5 & 0 & 5\\
    \textsf{NC\_STORAGE} & 36 & 53 & 89 \\
    \textsf{NC\_ICC} & 71 & 0 & 71\\
    \textsf{NC\_OUT\_STREAM} & 10 & 1 & 11 \\
    \textsf{NC\_OTHER} & 4 & 1 & 5 \\
  \hline
\end{tabular}
\vspace{-3ex}
\end{table}

The taint connection analysis can be further parameterized by different static data-flow analyses, e.g., \cite{DBLP:conf/ccs/WeiROR14, DBLP:conf/pldi/ArztRFBBKTOM14, DBLP:conf/ndss/GordonKPGNR15, DBLP:conf/eurosp/CalzavaraGM16}. In contrast to the tightly embedded annotations to label sources and sinks \cite{DBLP:conf/ndss/GordonKPGNR15}, the sources and sinks of \cite{DBLP:conf/ccs/WeiROR14, DBLP:conf/pldi/ArztRFBBKTOM14} are easier configured and more suitable to our analysis. Deciding the predefined sink list $S$ for the analysis in Algorithm~\ref{algo:dfa} is critical to define the impact of misuses. Specifically, we extend the default sinks of FlowDroid 2.7.1 (163 sinks) with 56 Android APIs involving network and serialization operations (e.g., the ones of \textsf{Parcel}/\textsf{Parcelable}, \textsf{ObjectOutputStream}, and \textsf{SQLiteDatabase}); see the number of sinks in Table~\ref{tab:sink-category}. Beyond this, we follow the categorization strategy of SuSi \cite{DBLP:conf/ndss/RasthoferAB14} to categorize the sinks for the efficiency of learning-based risk assessment. Moreover, for the sinks in \textsf{NO\_CATEGORY}(\textsf{NC}), we further divide these sinks into four sub-categories: \emph{ICC}, \emph{storage component}, \emph{output stream}, and \emph{others}. The ICC sinks include the APIs for inter-component communications. The sinks of storage components are the APIs of storage-related components, e.g., \textsf{SharedPreferences} and \textsf{ContentResolver}.

\subsection{Risk Assessment of Misuses}\label{subsec:clustering}

To assess the quantitative risk of cryptographic misuses, we assign severity weight $w_{id}$ to each vulnerability type in Table~\ref{tab:vul-weight}. The severity weights are compatible with CryptoGuard, and we only compensate the weights for $id=8,18\sim 21$. We also assign a risk weight $w_{sc}$ to each sink category, as explained in more detail in Section~\ref{sec:implementation}.

From the set of misuse tuples derived by Algorithm~\ref{algo:dfa}, we define two parameters for quantifying the risk of each app.
\begin{compactenum}
  \item The detectability $b_{\textsf{tool},i}$ predicates whether some misuse caused by the $i$-th vulnerability is detectable by the detector \textsf{tool}, where $\textsf{tool}\in \{\textsf{CG},\textsf{CC}\}$ for the minimal valid detector chain.
  \item $n_{sc,i}$ stands for the number of sensitive flows that the data affected by the $i$-th vulnerability can flow to a sink in category $sc$, where $sc\in\{\textsf{FILE},\textsf{LOG},\ldots\}$.
\end{compactenum}
When calculating $n_{sc,i}$, we removed the duplicates of the same misuse detected by multiple detectors. Consequently, $n_{sc,i}$ reflects the number of entry points on the attack surface of each app. We quantify the overall risk level of the individual app caused by the detected cryptographic misuses with a score $R_x$ that weights the severity of vulnerabilities and the risk of sink categories on the sensitive flows, i.e.,
\begin{displaymath}
R_x = \sum_{i=1}^{21} w_i\cdot \bigvee_{\textsf{tool}\in\{\textsf{CG},\textsf{CC}\}}b_{\textsf{tool},i} \cdot(\sum_{sc\in\{\textsf{FILE},\ldots\}} w_{sc}\cdot n_{sc,i})
\end{displaymath}

While the per-app risk assessment is useful for the development life cycle, our approach can also support app vetting at app stores. Due to the lack of ground truth of wild cryptographic misuses, we propose a clustering-guided risk prediction to capture the potential misuse threat patterns to large-scale apps. To predict the most significant misuse-caused threats for app clusters, we label each cluster with the top-ranking (\textit{vulnerability type}, \textit{sink category}) relations. Such top-ranking labels of each cluster compensate the app's risk value with significant knowledge of the tendency of misuse threats. In detail, the feature vector for clustering enumerates the orthogonal vulnerability types, sink categories, and tools. We get the total number of each type of data flow in one cluster and calculate the average number per app. A larger average number represents a broader attack surface on such data flows. We use the $(id,sc)$ pair of the largest average numbers to label the cluster and predict the most significant threats to the apps of this cluster. The top-label-based prediction is adept at presenting an overall threat summary of each cluster than only counting the largest numbers of sensitive flows from the feature vector of each app. As a result, we reach effective risk predictions for app vetting.

\begin{table}[t]
\renewcommand{\arraystretch}{1.1}
  \caption{Severity Weight of Vulnerability}\small
  \label{tab:vul-weight}
  \centering
  \begin{tabular}{cc}
    \hline
    Vulnerability ID ($id$) & Severity Weight ($w_{id}$)\\
    \hline
    1$\sim$8, 17 & high (10)\\
    9$\sim$13, 19 & medium (7)\\
    14$\sim$16, 18 & low (4)\\
    20, 21 & very low (1) \\
  \hline
\end{tabular}
\vspace{-2ex}
\end{table}

\section{Implementation Settings}\label{sec:implementation}

With the framework proposed in Section~\ref{sec:approach}, our implementation can be parameterized at least on the following aspects: 1) the valid detector chains for the misuse detection, 2) the choice of sinks and the risk weight assignment $w_{sc}$ for each sink category, 3) the static data-flow analyses for Algorithm~\ref{algo:dfa}, and 4) the clustering algorithms. To develop an instance of the framework to demonstrate our evaluations, we used FlowDroid \cite{DBLP:conf/pldi/ArztRFBBKTOM14} as the taint-connection analysis of the misuse-originating data-flow analysis and developed $DSTrack$ with Androguard \cite{androguard}. Unlike the data-flow analysis used in identifying the misuse patterns \cite{DBLP:conf/ccs/RahamanXASTFKY19,cognicrypt}, our data-flow analysis is mainly used to capture the dependencies of data leakage to the cryptographic misuses, and $DSTrack$ takes an ad-hoc approach backward through the control-flow paths. $DSTrack$ is intra-procedural, while the taint-connection analysis with FlowDroid is inter-procedural, context-, flow-, and object-sensitive. Then, we use $k$-means clustering \cite{macqueen1967some} for Euclidean distance to group the apps. Next, we discuss the choices for the first two parameters in detail.

Our framework is extensible to adapt more state-of-the-art detectors. Generally, when adding or removing a detector, we should preserve Corollary~\ref{corollary1} of a valid detector chain, which is the main criterion for selecting detectors used in \sysname{}. Even if the misuse rules addressed by one detector are only a subset of the vulnerabilities of another detector, adapting more detectors into the detector chain is still meaningful. First, a new detector may detect more vulnerability sites and sensitive flows. Second, the new adapter provides more features for the clustering-guided assessment, which affect the top-ranking labels of each cluster. We have developed three adapters to assemble CryptoGuard, \textsc{CogniCrypt}$_\text{SAST}$, and BinSight \cite{DBLP:conf/ccs/MuslukhovBB18} into the detector chain. The adapters implement a unified interface for extracting detector output to build the misuse tuples, converting misuse information from \textsf{smali} to Java, and mapping each misuse rule to one vulnerability $id$ in our threat model. Our implementation consists of 3.2 kLoC of Java and Python, including adapters, data-flow analysis, feature extraction, learning, and glue code.

Among all the parameters, the most flexible one is the choice of sinks and the assignment of their risk weights since the sinks may be derived and classified into different categories by different strategies \cite{DBLP:conf/ndss/RasthoferAB14, DBLP:conf/trust/GiblerCEC12}.
Different categorizations of sinks lead to different assessment results because they may affect the dimension of the feature vector and the number of sensitive flows falling into each category. The sinks of our implementation are an extension of the default sinks of FlowDroid. The rationality of risk weights for the sink category is another issue impacting the quantitative risk of each app. To decide the risk weights $w_{sc}$, we use the keywords in the signature of sink APIs to search for the Android-relevant CVEs and the corresponding CVSS metrics in NVD. We find a partial order of the sink categories based on the average value of the CVSS scores correlated to the sinks of each sink category. Then we normalize and approximate the risk ranking to get the values in Table~\ref{tab:sink-weight}. While the risk ranking of sinks is important for per-app assessment, we choose to use the detectability by detectors ($b_{\textsf{tool},i}$) and the number of sensitive flows ($n_{sc,i}$) to decide the features for clustering. This choice can avoid the threat summary of each cluster being affected by the potential bias of risk weights.

\begin{table}[!t]
\renewcommand{\arraystretch}{1.1}
  \caption{Risk Weight of Sink Category}\small
  \label{tab:sink-weight}
  \centering
  \begin{tabular}{lc}
    \hline
    Sink Category ($sc$) & Risk Weight ($w_{sc}$)\\
    \hline
    \textsf{SMS\_MMS}, \textsf{NC\_OTHER} & 1 \\
    \textsf{LOG} & 3 \\
    \textsf{SYNC}, \textsf{NC\_STORAGE} & 4\\
    \textsf{FILE}, \textsf{NC\_OUT\_STREAM} & 5\\
    \textsf{NC\_ICC} & 7\\
    \textsf{NETWORK} & 10\\
  \hline
\end{tabular}
\vspace{-2ex}
\end{table}

\section{Evaluation}\label{sec:evaluation}
In this section, we investigate the accuracy of \sysname{} and the effect of risk assessment over the flow-related features. Moreover, we make useful security observations based on the results of an empirical study and the analysis of popular real-world apps.

We evaluate \sysname{} on three app datasets, as shown in Table~\ref{tab:dataset}. We use dataset S2 to evaluate the accuracy of \sysname{} and its ability to reveal threats on the attack surface of real-world apps. CryptoAPI-Bench in S2 is a public crypto-misuse benchmark with ground truths to evaluate accuracy. We also take another evaluation on 15 popular real-world apps to estimate the expected true positives and the precision improvement by augmenting the detector chain from another perspective. We use dataset S1 to evaluate the comprehensiveness of our vulnerability list and the effect of risk assessments on app store-side vetting. Dataset S3 consists of new releases of the 15 real-world apps in S2, which we used to discuss the vulnerability fixing related to apps' evolution.

Our experiments are conducted on an elastic compute service with 2.5GHz$\times$ 4 Intel Xeon(R) Platinum 8269CY CPU, 16GB RAM, Linux 4.15.0-88-generic kernel (Ubuntu 18.04), and JDK 1.8. For the detector chain, we use CryptoGuard 03.04.00 (commit id 1d520e4) \cite{DBLP:conf/ccs/RahamanXASTFKY19}, \textsc{CogniCrypt}$_\textsc{sast}$-Android 1.0.0 (commit id 856b1da) \cite{cognicrypt-android}, and BinSight (commit id cd8b680) \cite{DBLP:conf/ccs/MuslukhovBB18}. For the accuracy investigation, we also use \textsc{CogniCrypt}$_\textsc{sast}$ 2.7.1 (commit id 98eccd4) for Java \cite{DBLP:conf/ecoop/KrugerS0BM18, cognicrypt}. For data-flow analysis, we use FlowDroid 2.7.1 (commit id 72734bd) \cite{DBLP:conf/pldi/ArztRFBBKTOM14}. For the data-source tracking ($DSTrack$), we use Androguard (commit id 22849b6) \cite{androguard}. We aim to answer the following research questions.

\begin{compactenum}[RQ1.]
\item What is the accuracy of the valid detector chains of \sysname{}?
\item Is our comprehensive vulnerability list better than the existing vulnerability types and rules to specify the misuse?
\item What can we find on the attack surface of real-world apps with our framework?
\item How can the risk assessment of cryptographic misuses benefit the app store-side vetting?
\end{compactenum}

\begin{table}[t]
\renewcommand{\arraystretch}{1.1}
  \caption{Summary of Datasets}\scriptsize
  \label{tab:dataset}
  \begin{tabular}{crl}
    \hline
    Dataset & \#App & Description\\
    \hline
    S1 & 40604 & Released Jan.$\sim$Dec. 2019 on Google Play, Anzhi, and other\\
    & &   popular app markets, and downloaded through AndroZoo \cite{Allix:2016:ACM:2901739.2903508} \\
    S2 & 190 & 175 from CryptoAPI-Bench (commit id 759622f) \cite{DBLP:conf/secdev/AfroseRY19}, 15 \\
    & & popular real-world apps (Chrome, PayPal, etc) released in 2020 \\

    S3 & 15 & The new releases of the 15 real-world apps in S2 (released \\
    & & from Jun. to Oct. 2022)\\
  \hline
\end{tabular}
\vspace{-2ex}
\end{table}

\subsection{Accuracy of Detection (RQ1)}

\begin{table*}[t]
\renewcommand{\arraystretch}{1.1}
\centering
  \caption{Severity-classified metrics comparison of \sysname{}, CryptoGuard and \textsc{CogniCrypt}$_\text{SAST}$. TP/FP/FN are the number of true positives/false positives/false negatives respectively}
  \label{tab:severity-metrics}
  \scriptsize
  \begin{tabular}{c|c|ccc|ccc|ccc|ccc}
    \hline
    Severity & & \multicolumn{3}{c|}{CryptoGuard} & \multicolumn{3}{c|}{\textsc{CogniCrypt}$_\text{SAST}$} & \multicolumn{3}{c|}{\sysname{}} & \multicolumn{3}{c}{\sysname{}} \\
    Weight & GTP & \multicolumn{3}{c|}{(\textsf{CG})} & \multicolumn{3}{c|}{(\textsf{CC})} & \multicolumn{3}{c|}{(\textsf{CG}+\textsf{CC})} & \multicolumn{3}{c}{(\textsf{CG}+\textsf{CC}+\textsf{BS})} \\\cline{3-14}
    ($w_{id}$) & & TP & FP & FN & TP & FP & FN & TP & FP & FN & TP & FP & FN \\
    \hline
    1 & 67 & 0 & 0 & 67 & 67 & 0 & 0 & 67 & 0 & 0 & 67 & 0 & 0 \\
    4 & 42 & 39 & 7 & 3 & 13 & 7 & 29 & 40 & 7 & 2 & 40 & 7 & 2 \\
    7 & 36 & 32 & 5 & 4 & 14 & 4 & 22 & 34 & 5 & 2 & 35 & 5 & 1 \\
    10 & 63 & 57 & 8 & 6 & 26 & 10 & 37 & 63 & 11 & 0 & 63 & 11 & 0 \\
    \hline
    Total & 208 & 128 & 20 & 80 & 120 & 21 & 88 & 204 & 23 & 4 & 205 & 23 & 3\\
    \hline
    \multicolumn{2}{c|}{Precision(\%)} & \multicolumn{3}{c|}{86.5} & \multicolumn{3}{c|}{85.1} & \multicolumn{3}{c|}{89.9} & \multicolumn{3}{c}{89.9}\\
    \multicolumn{2}{c|}{Recall(\%)} & \multicolumn{3}{c|}{61.5} & \multicolumn{3}{c|}{57.7} & \multicolumn{3}{c|}{98.1} & \multicolumn{3}{c}{98.6} \\
    \multicolumn{2}{c|}{F1(\%)} & \multicolumn{3}{c|}{71.9} & \multicolumn{3}{c|}{68.8} & \multicolumn{3}{c|}{93.8} & \multicolumn{3}{c}{94.0}\\
    \hline
  \end{tabular}
\vspace{-2ex}
\end{table*}

To evaluate the accuracy of the adapter-based detection of \sysname{}, firstly, we compare the minimal valid detector chain (\textsf{CG}+\textsf{CC}) with the standalone CryptoGuard and \textsc{CogniCrypt}$_\text{SAST}$ on the CryptoAPI-Bench of dataset S2. The statistics and comparison of metrics (precision, recall, and F1) are presented in Table~\ref{tab:severity-metrics}. Because we are dealing with a superset of vulnerability types compared with CryptoAPI-Bench, we manually refine the original ground truth of CryptoAPI-Bench (in Table VI,\cite{DBLP:conf/secdev/AfroseRY19}) to build the ground truth w.r.t. the new comprehensive vulnerability list, i.e., the 208 ground truth positives (GTPs) in Table~\ref{tab:severity-metrics}. Then the true positives (TP), false positives (FP), and false negatives (FN) of different approaches are identified by checking the detection results against the GTPs. We summarize the numbers of TP/FP/FN on different severity levels according to the severity weights of vulnerability types. When addressing our new threat types, the test cases introduce 67 incomplete usages of cryptographic API (the 67 misuses of the vulnerability type 20), which \textsc{CogniCrypt}$_\text{SAST}$ can detect. Assembling multiple detectors into \sysname{}, though tending to introduce more false positives (FP$_{\textsf{CG+CC}}$= FP$_{\textsf{CG}} \cup$FP$_{\textsf{CC}}$), can significantly suppress the false negatives (FN$_{\textsf{CG+CC}}$=(FN$_{\textsf{CG}}$-TP$_{\textsf{CC}}$)$\cup$(FN$_{\textsf{CC}}$-TP$_{\textsf{CG}}$)) and improve detection recall. Since we also merge the true positives, the precision can still be improved with more detectors. Then, we introduce a third detector, i.e., BinSight (\textsf{BS}), over the minimal detector chain (\textsf{CG}+\textsf{CC}). After adding BinSight, we found that the recall and F1 of the new detector chain (\textsf{CG}+\textsf{CC}+\textsf{BS}) increased compared with the case when only \textsf{CG} and \textsf{CC} were in the detector chain. Intuitively, introducing more detectors for misuse detection will also restrain the false negatives but potentially introduce new false positives. The improvement from BinSight is minor since BinSight only detects the positives of vulnerability type 1,9,11$\sim$14.

The metrics based on ground truths of CryptoAPI-Bench do not help decide the proper sources for the misuse-originating data-flow analysis of real-world apps because we have no ground truths for real-world apps, given the absence of source code. We propose a voting mechanism to estimate the \emph{expected true positives}. The principle is, among all the misuses detected by the detectors (i.e., positives), we determine a misuse to be an expected true positive only if $>$50\% of detectors capable of dealing with the corresponding vulnerability type report it as a misuse. Then, the precision of the detector chain is decided by its detection results against the expected true positives. The voting leads to a fair metric for estimating the true positives without human aid, assuming that each detector in the detector chain is independent. Relying on this voting mechanism, we evaluate the 15 popular real-world apps in S2 (their names listed in Table~\ref{tab:real-world-findings}) to derive the expected true positives and demonstrate the effect of augmenting the detector chain. We found that the precision of the detector chain (\textsf{CG}+\textsf{CC}+\textsf{BS}) decided by voting increases from 77.6\% to 78.7\% compared with the case that only \textsf{CG} and \textsf{CC} are in the detector chain. Moreover,
the expected true positives derived by the three-detector chain are confirmed manually and used in the evaluation of RQ3.

\vspace{1ex}
\noindent \framebox[\linewidth][l]{
\parbox{3.3in}{
The detector chain of \sysname{} outperforms the individual detectors on precision and recall based on CryptoAPI-Bench. The expected true positives voted by the detectors of the detector chain serve as a proper source of data-flow-based risk assessment.
}
}

\subsection{Comprehensiveness of Vulnerability List (RQ2)}

\begin{figure}[t]
  \centering
  \includegraphics[width=\linewidth]{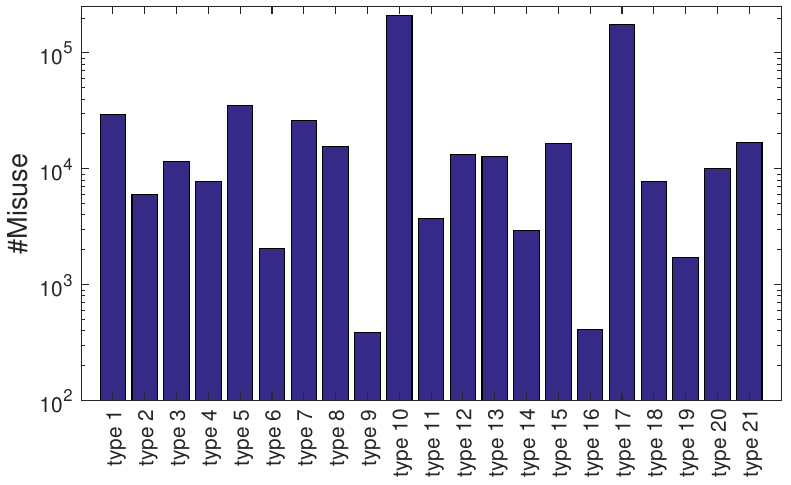}
  \caption{Distribution of cryptographic misuses on different vulnerability types}\label{fig:misuse}
\vspace{-2ex}
\end{figure}

We performed the adapter-based misuse detection of \sysname{} with the minimal valid detector chain (\textsf{CG}+\textsf{CC}) on dataset S1. We found there are 33410 apps detected to have some cryptographic misuse. Among these apps, CryptoGuard reports misuses on 30466 apps, and \textsc{CogniCrypt}$_\text{SAST}$ reports misuses on 19626 apps. The distribution of misuses on each vulnerability type is presented in Figure~\ref{fig:misuse}. The insecure PRNGs (type 10, reported 211283 cases) and insecure cryptographic hash operations (type 17, reported 175450 cases) are the most prominent vulnerabilities in the apps. The predictable/constant PRNG seeds (type 9, reported 389 cases) and insecure asymmetric ciphers (type 16, reported 411 cases) rarely appear. The SSL/TLS-related vulnerabilities (type 4$\sim$8) involve 86693 cases, accounting for 14.3\% of the detected misuses. From the statistics, we found that with intensive attention to the traditional vulnerability types (type 1,9,11$\sim$14), such misuses have become less pervasive (only 10.3\%) in real-world apps. If we compare with recent threat models, the threat model of CryptoGuard \cite{DBLP:conf/ccs/RahamanXASTFKY19} (lacking vulnerability type 8,18$\sim$21) will miss 8.6\% of cryptographic misuses. The rules of \textsc{CogniCrypt}$_\text{SAST}$ \cite{DBLP:conf/ecoop/KrugerS0BM18} (lacking vulnerability type 4$\sim$7) will miss 11.7\% of cryptographic misuses. Therefore, the threat models proposed by such approaches will cause considerable missing on important vulnerability types.

\vspace{1ex}
\noindent \framebox[\linewidth][l]{
\parbox{3.3in}{
The insecure PRNGs (type 10) and insecure cryptographic hash operations (type 17) are the most prominent vulnerabilities. Traditional misuses (type 1,9,11$\sim$14) have become less pervasive (10.3\%), and more recent approaches \cite{DBLP:conf/ecoop/KrugerS0BM18, DBLP:conf/ccs/RahamanXASTFKY19} still miss a considerable number of misuses, which validates our direction to synthesize a more comprehensive threat model.
}
}

\subsection{Effect of Data-Flow-Driven Risk Assessment and Security Findings (RQ3)}

\begin{figure}[t]
  \centering
  \includegraphics[width=\linewidth]{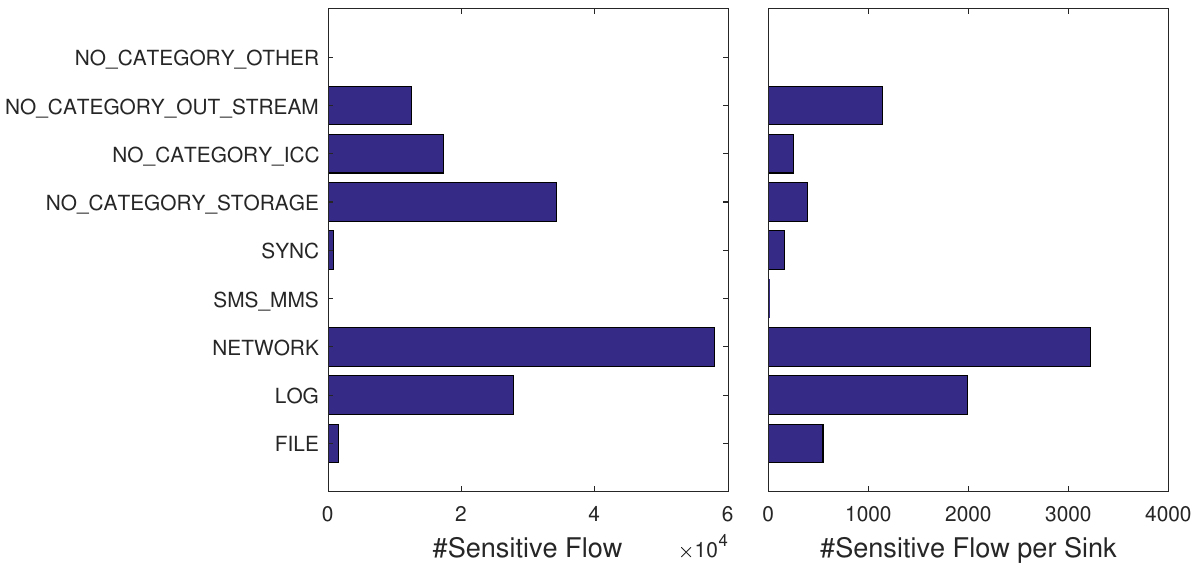}
  \caption{Distribution of data flows triggered by cryptographic misuses on different sink categories}\label{fig:sink}
\vspace{-2ex}
\end{figure}

We conducted the data-flow analysis of \sysname{} on the 33410 apps with cryptographic misuses and found 24055 apps (72.0\%) containing data flows originating from cryptographic misuse. On the misuse level, we found 152305 misuse-originating sensitive data flows. We classify these sensitive data flows into different sink categories, as shown in Figure~\ref{fig:sink}. We observed that for the two most risky sink categories, there are 57909 (38.0\%) sensitive flows to \textsf{NETWORK} sinks and 17347 (11.4\%) sensitive flows to \textsf{NC\_ICC} sinks. Meanwhile, there are only 2 sensitive flows to \textsf{SMS\_MMS} sinks, and none flows to \textsf{NC\_OTHER} sinks from cryptographic misuses. For each sink method in different sink categories, we find that the \textsf{NETWORK} sinks and \textsf{LOG} sinks are the top 2 common entry points on the attack surface to exploit the vulnerabilities caused by cryptographic misuses.

\begin{table*}
  \caption{Threats on Real-world Apps identified by \sysname{}. \textit{other} stands for the misuses with vulnerability types which trigger no sensitive flows on these apps. Sink categories abbreviated, e.g. \text{O\_S} = \text{NC\_OUT\_STREAM}}
  \label{tab:real-world-findings}
  \scriptsize
  \begin{tabular}{p{1.2cm}|p{0.6cm}|p{1cm}|p{1cm}|p{1cm}|p{0.6cm}|p{1cm}|p{0.6cm}|p{1.1cm}|p{1cm}|p{1cm}|p{1cm}|p{0.4cm}|r}
    \hline
    & \multicolumn{12}{c|}{Vulnerability $id$} & \\ \cline{2-13}
    App & 1 & 3 & 8 & 10 & 12 & 13 & 15 & 17 & 18 & 20 & 21 & other & $R_x$ \\
    \hline

    Amazon Shopping & 5 (\textsf{STO}:2) & 2 & 2 & 13 & 2 (\textsf{STO}:4) & & 1 (\textsf{STO}:4) & 9 (\textsf{STO}:1) & & 6 (\textsf{NET}:2, \textsf{STO}:24, \textsf{SYNC}:3) & 1 & 6 & 424 \\\hline

    Chrome & 2 & & & 11 & & & & 10 & & 5 & & 1 & 0 \\\hline

    Booking & 1 & 1 (\textsf{NET}:1) & 1 (\textsf{NET}:1) & 7 & & 6 (\textsf{O\_S}:6) & 1 & 3 & 1 (\textsf{O\_S}:1) & 2 (\textsf{O\_S}:1) & & 22 & 435 \\\hline

    Moneycontrol & & & & 7 & & & & 14 & & 1 & & 1 & 0 \\\hline

    eBay & 24 & 1 (\textsf{O\_S}:1, \textsf{NET}:1) & 1 (\textsf{O\_S}:1, \textsf{NET}:1) & 3 & & & & & & & & 18 & 300 \\\hline

    PayPal & 1 & 1 & 4 & 4 & & & 2 & 7 & 1 & 1 & 2 & 1 & 0 \\\hline

    Uber & 1 & 1 (\textsf{ICC}:1) & 7 (\textsf{NET}:2, \textsf{O\_S}:1, \textsf{ICC}:5) & 1 & & & & 6 & & & 2 (\textsf{ICC}:1) & 2 & 677 \\\hline

    Lyft & & 1 & 3 (\textsf{NET}:2, \textsf{O\_S}:1) & 9 & & & & 7 & & & 1 & & 250 \\\hline

    Facebook Lite & & 2 (\textsf{ICC}:1) & 5 (\textsf{ICC}:3) & 11 & & & & 9 & 1 & & 5 (\textsf{ICC}:3) & 3 & 301 \\\hline

    WeChat & 7 & & & 13 & 1 & 1 & 7 & 24 & 2 & 1 & 2 & 2 & 0 \\\hline

    TikTok & 2 & & & 11 & 2 & & & 9 (\textsf{ICC}:2) & & & & 1 & 140 \\\hline

    PNC Mobile & 2 & 3 (\textsf{LOG}:1) & 3 & 4 & 1 & & & 6 & & & 4 (\textsf{LOG}:1) & 2 & 33\\\hline

    CIBC Mobile Banking & 1 & 1 (\textsf{STO}:1, \textsf{NET}:6, \textsf{LOG}:2, \textsf{FILE}:1) & 3 (\textsf{STO}:2, \textsf{NET}:13, \textsf{LOG}:4, \textsf{FILE}:2) & 5 & 1 (\textsf{ICC}:3, \textsf{O\_S}:1) & & & 10 (\textsf{STO}:3, \textsf{NET}:2, \textsf{LOG}:2, \textsf{O\_S}:1, \textsf{ICC}:1) & 1 & 1 & 3 (\textsf{STO}:1, \textsf{NET}:6, \textsf{LOG}:2, \textsf{FILE}:1) & 4 & 3107 \\\hline

    Starbucks & 2 & 2 (\textsf{STO}:2) & 4 (\textsf{STO}:2, \textsf{O\_S}:1) & & & 1 & & 1 & & & 4 (\textsf{STO}:6, \textsf{O\_S}:1) & & 239 \\\hline

    Walmart & & 1 (\textsf{NET}:2, \textsf{LOG}:1) & 2 & 8 (\textsf{NET}:4, \textsf{LOG}:2) & & & & 17 (\textsf{STO}:1) & & & 1 & & 730 \\
    \hline
  \end{tabular}
\vspace{-2ex}
\end{table*}

To show our framework's ability to discover threats in real-world apps, we investigate the vulnerabilities caused by cryptographic misuses and the related sensitive flows in the 15 popular real-world apps of dataset S2. The results are presented in Table~\ref{tab:real-world-findings}. In this evaluation, the expected true positives derived by the detector chain have to be confirmed manually to ensure an effective attack surface estimation. For each app, we count the expected true positives of misuses identified by the valid detector chain (\textsf{CG}+\textsf{CC}+\textsf{BS}). Note that \textsf{BS} is adapted only to decide the expected true positive misuses. We do not merge it into the definition of $R_x$ in Section~\ref{subsec:clustering}. Then we report the sensitive flows originating from these misuses in the brackets in Table~\ref{tab:real-world-findings} in terms of \textit{sink\_category:num\_flows}. We also calculate the risk value $R_x$ for each app. We figure out that among all risky misuses ($w_{id}=7,10$), type 3,8,17 are more likely to trigger sensitive data flows. Moreover, the misuses of type 3 and 8 are likely to trigger flows to more risky sinks (\textsf{NETWORK} and \textsf{NC\_ICC}). It means that in typical apps, vulnerable passwords for KeyStore (type 3) and usage of the expired protocol by SSLContext (type 8) are more likely to be responsible for the data leakages caused by the misuses, which should be paid close attention to in developing apps. Also, we have found risky apps (with bigger $R_x$), including a financial app \emph{CIBC mobile banking}. We manually confirmed that the threat caused by the type 8 misuses detected in this app has also been reported by CVE-2014-5594 \cite{CVE-2014-5594}.

It is interesting to investigate how the threats to the real-world apps vary along with their evolution, as presented in Table~\ref{tab:migration}. The dataset S3 contains the new releases of the 15 real-world apps. Both the number of misuses and the risk score $R_x$ get suppressed in the newer version of the 8 apps. Especially, the risk of \emph{CIBC mobile banking}, seriously threatened by the type-8 misuse, has drastically reduced. \emph{Walmart} app introduces more misuses but has a lower risk in its new version. The \emph{Moneycontrol} and \emph{Facebook Lite} app, though whose newer releases have fewer misuses detected, have been exposed to new entry points on the attack surface. The new \emph{Moneycontrol} has new type-3 and type-8 misuses leading to \textsf{NETWORK} flows. The new \emph{Facebook Lite} has more type-3 and type-8 misuses leading to either \textsf{NETWORK} and \textsf{OUT\_STREAM}. Only the \emph{Starbucks} app has both increased misuses and risk score in its new version. We manually investigated the risk increases in the 3 apps. The risks are caused by calling new sinks from the app code to catch the output of the misused APIs. \emph{Starbucks} and \emph{Moneycontrol} also use new third-party libraries. However, these third-party libraries are annotation libraries, e.g., \textsf{jsr305}, bringing no sensitive data flows into the app.

\begin{table}
\renewcommand{\arraystretch}{1.2}
  \caption{Threats Variation of the 15 Real-World Apps}
  \label{tab:migration}
  \scriptsize
  \centering
  \begin{tabular}{l|r|r|r|r}
    \hline
     & \multicolumn{2}{c|}{\#Vulnerability} & \multicolumn{2}{c}{$R_x$} \\ \cline{2-5}
    App & S2 & S3 & S2 & S3 \\
    \hline
    Amazon Shopping & 47 & 9 & 424 & 0 \\

    Chrome & 29 & 11 & 0 & 0 \\

    Booking & 45 & 2 & 435 & 0 \\

    Moneycontrol & 23 & 11 & 0 & 600 \\

    eBay & 47 & 3 & 300 & 0 \\

    PayPal & 24 & 28 & 0 & 0 \\

    Uber & 20 & 6 & 677 & 0 \\

    Lyft & 21 & 3 & 250 & 0\\

    Facebook Lite & 36 & 25 & 301 & 1424\\

    WeChat & 60 & 23 & 0 & 0\\

    TikTok & 25 & 13 & 140 & 0 \\

    PNC Mobile & 25 & 18 & 33 & 0 \\

    CIBC Mobile Banking & 30 & 12 & 3107 & 130 \\

    Starbucks & 14 & 20 & 239 & 710 \\

    Walmart & 29 & 51 & 730 & 300 \\
  \hline
\end{tabular}
\vspace{-3ex}
\end{table}

\vspace{1ex}
\noindent \framebox[\linewidth][l]{
\parbox{3.3in}{
The \textsf{NETWORK} and \textsf{LOG} sinks are the most common access points on the attack surface to exploit the vulnerabilities caused by the misuses. The app's quantitative risk value $R_x$ effectively assesses the risk of cryptographic misuses. Our per-app assessment reports a severe real-world threat in a financial app confirmed by a CVE record.
}
}

\subsection{Effect of Risk Assessments to App Vetting (RQ4)}

To show the effect of our framework in the app-vetting process for app stores, we apply the clustering-based assessment on the 33410 apps that contain cryptographic misuses. We label each cluster with the most frequently occurring (\textit{vul\_type}, \textit{sink\_category}) pairs. We consider both the Davies-Bouldin Index (DBI) \cite{4766909} and the discrimination of top-ranking labels of each cluster to decide $k$, the proper number of clusters. On dataset S1, we found a proper prediction model when $k=7$. The top-3 labels of each cluster are presented in Table~\ref{tab:clustering}. In cluster $c_1$, 53.1\% of detected sensitive flows are from customized \textsf{TrustManager} (vulnerability type 5) to \textsf{NETWORK} sink. All the apps in this cluster contain at least one sensitive flow with this label, and the average number of this kind of flow is 2.42 in each app. The small clusters $c_2$ and $c_5$ address the impact of static initialization vectors in CBC-mode ciphers (type 13) on inter-component communications. The difference is that in $c_2$, the static IVs also affect the storage components remarkably, while in $c_5$, the impacts are more from insecure 64-bit block ciphers. Around 50\% of sensitive flows in $c_3$ result in leakages to log, either from 64-bit block ciphers (type 15) or insecure keys (type 1). Over 95\% of sensitive flows in $c_4$ lead to the storage components, mainly from suspected usages under more testing (type 21) and vulnerable passwords for \textsf{KeyStore} (type 3). The insecure cryptographic hash operations (type 17) are in the lead of the source of sensitive flows in cluster $c_6$ and $c_7$. In detail, the insecure hash operations trigger over 23\% of sensitive flows in $c_6$ and over 32\% in $c_7$. We also found that over 30\% of sensitive flows in $c_6$ affect the storage components from the insecure hash, static IVs, and constant keys. Moreover, all the apps with no sensitive flows (33410 - 24055 = 9355) are grouped into $c_7$. However, the amount of top-ranking sensitive flows in $c_6$ and $c_7$ is far from dominating all detected flows in these clusters, which indicates a richer diversity of sensitive flows in these clusters.

\begin{table}[t]
\renewcommand{\arraystretch}{1.1}
  \caption{Clustering results on Dataset S1. \#Labels per app = (\#specific flows)/(\#apps in each cluster). \% of labels = (\#specific flows)/(\#all detected flows in each cluster)}
  \label{tab:clustering}
  \scriptsize
  \centering
  \begin{tabular}{c|r|c|r|r|r}
    \hline
    Cluster & \#App & Label($id\rightsquigarrow sc$) & \#App & \#Label per & \% of \\
    ID & & & with label & app (avg.) & labels \\
    \hline

    & & 5$\rightsquigarrow$\textsf{NETWORK} & 13821 & 2.42 & 53.1 \\\cline{3-6}
    $c_1$ & 13821 & 4$\rightsquigarrow$\textsf{NETWORK} & 4606 & 0.46 & 10.2 \\\cline{3-6}
    & & 8$\rightsquigarrow$\textsf{NETWORK} & 859 & 0.19 & 4.1 \\\hline

    & & 13$\rightsquigarrow$\textsf{NC\_ICC} & 26 & 210.85 & 98.2 \\\cline{3-6}
    $c_2$ & 26 & 13$\rightsquigarrow$\textsf{NC\_STORAGE} & 25 & 3.35 & 1.6 \\\cline{3-6}
    & & 13$\rightsquigarrow$\textsf{LOG} & 12 & 0.58 & 0.3 \\\hline

    & & 15$\rightsquigarrow$\textsf{LOG} & 1531 & 4.01 & 31.7 \\\cline{3-6}
    $c_3$ & 1542 & 1$\rightsquigarrow$\textsf{LOG} & 900 & 2.30 & 18.2 \\\cline{3-6}
    & & 15$\rightsquigarrow$\textsf{NC\_STORAGE} & 1061 & 1.47 & 11.6 \\\hline

    & & 21$\rightsquigarrow$\textsf{NC\_STORAGE} & 1668 & 3.69 & 47.5 \\\cline{3-6}
    $c_4$ & 1671 & 3$\rightsquigarrow$\textsf{NC\_STORAGE} & 1649 & 3.63 & 46.8 \\\cline{3-6}
    & & 8$\rightsquigarrow$\textsf{NC\_STORAGE} & 25 & 0.09 & 1.1 \\\hline

    & & 13$\rightsquigarrow$\textsf{NC\_ICC} & 16 & 90.00 & 86.9 \\\cline{3-6}
    $c_5$ & 16 & 15$\rightsquigarrow$\textsf{NC\_ICC} & 8 & 4.50 & 4.3 \\\cline{3-6}
    & & 8$\rightsquigarrow$\textsf{NETWORK} & 10 & 1.38 & 1.3 \\\hline

    & & 17$\rightsquigarrow$\textsf{NC\_STORAGE} & 335 & 5.50 & 16.6 \\\cline{3-6}
    $c_6$ & 401 & 7$\rightsquigarrow$\textsf{NETWORK} & 96 & 2.68 & 8.1 \\\cline{3-6}
    & & 13$\rightsquigarrow$\textsf{NC\_STORAGE} & 194 & 2.62 & 7.9 \\\hline

    & & 17$\rightsquigarrow$\textsf{LOG} & 1854 & 0.27 & 11.9 \\\cline{3-6}
    $c_7$ & 15933 & 17$\rightsquigarrow$\textsf{NC\_ICC} & 1260 & 0.24 & 10.7 \\\cline{3-6}
    & & 20$\rightsquigarrow$\textsf{LOG} & 413 & 0.16 & 7.0 \\\hline

  \end{tabular}
\vspace{-2ex}
\end{table}

One interesting discovery is that the insecure PRNGs (type 10), though detected most frequently (see Figure~\ref{fig:misuse}), are in general unlikely to trigger sensitive flow, which means more occurrences of the misuses do not confirm a broad attack surface. Besides, an app falling into a specific cluster will be predicted to have a higher chance of triggering some specific kinds of sensitive flows. Moreover, the clustering result gives a more detailed summary of the most representative threats found in each cluster. This kind of summary can further guide the app-store-side vetting strategies. For example, some hybrid analysis \cite{DBLP:conf/ndss/SounthirarajSGLK14} on MITM attacks may be applied with priority on the apps in cluster $c_1, c_5$, and $c_6$ because the SSL/TLS-related vulnerabilities (type 4$\sim$8) tend to trigger sensitive flows to the network in these clusters.

\begin{figure}[t]
  \centering
  \includegraphics[width=\linewidth]{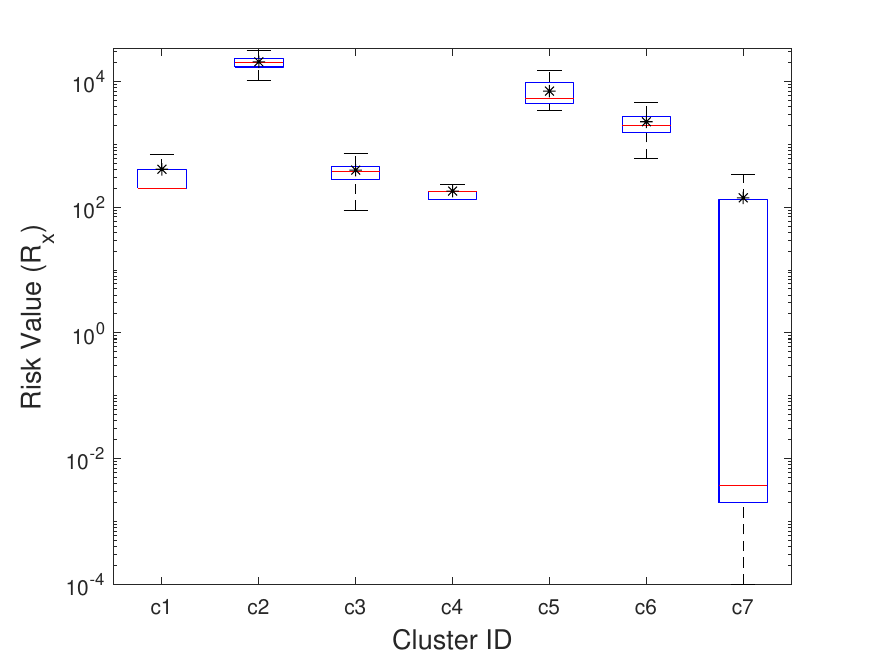}
  \caption{Quantitative risk levels of each cluster}\label{fig:quantitative-risk}
\vspace{-2ex}
\end{figure}

\begin{table}[!t]
\renewcommand{\arraystretch}{1.1}
  \caption{Association rules generated with FP-growth, selected by (\#app instance $>500$) and (confidence $>0.8$). Gray rules support associations between top-ranking labels in Table~\ref{tab:clustering}}
  \label{tab:association}
  \footnotesize
  \centering
  \begin{tabular}{l|r}
    \hline
    Association rule & Conf \\
    \hline
    $[3\rightsquigarrow\textsf{NETWORK}]$:1089 $\Rightarrow [8\rightsquigarrow \textsf{NETWORK}]$:1076 & 0.99 \\
    \cellcolor{lightgray}$[21\rightsquigarrow\textsf{NC\_STORAGE}]$:1929 $\Rightarrow [3\rightsquigarrow \textsf{NC\_STORAGE}]$:1769 & 0.92 \\
    $[15\rightsquigarrow\textsf{NC\_OUT\_STREAM}]$:1300 $\Rightarrow [15\rightsquigarrow\textsf{LOG}]$:1186 & 0.91 \\
    $[8\rightsquigarrow\textsf{NC\_STORAGE}]$:1002 $\Rightarrow [5\rightsquigarrow \textsf{NETWORK}]$:903 & 0.90 \\
    $[15\rightsquigarrow\textsf{NC\_OUT\_STREAM}]$:1300 $\Rightarrow [15\rightsquigarrow \textsf{NC\_STORAGE}]$:1161 & 0.89 \\
    \cellcolor{lightgray}$[8\rightsquigarrow\textsf{NC\_STORAGE}]$:1002 $\Rightarrow [3\rightsquigarrow \textsf{NC\_STORAGE}]$:884 & 0.88 \\
    \cellcolor{lightgray}$[15\rightsquigarrow\textsf{NC\_STORAGE}]$:1387 $\Rightarrow [15\rightsquigarrow \textsf{LOG}]$:1204 & 0.87 \\
    $[18\rightsquigarrow\textsf{NC\_STORAGE}]$:1021 $\Rightarrow [17\rightsquigarrow \textsf{LOG}]$:868 & 0.85 \\
    \cellcolor{lightgray}$[13\rightsquigarrow\textsf{LOG}]$:687 $\Rightarrow [13\rightsquigarrow\textsf{NC\_STORAGE}]$:582 & 0.85 \\
    \cellcolor{lightgray}$[1\rightsquigarrow\textsf{LOG}]$:1277 $\Rightarrow [15\rightsquigarrow\textsf{LOG}]$:1070 & 0.84 \\
    $[15\rightsquigarrow\textsf{NC\_STORAGE}]$:1387 $\Rightarrow [15\rightsquigarrow \textsf{NC\_OUT\_STREAM}]$:1161 & 0.84 \\
    $[13\rightsquigarrow\textsf{LOG}]$:687 $\Rightarrow [15\rightsquigarrow\textsf{LOG}]$:572 & 0.83 \\
    \cellcolor{lightgray}$[4\rightsquigarrow\textsf{NETWORK}]$:5832 $\Rightarrow [5\rightsquigarrow\textsf{NETWORK}]$:4826 & 0.83 \\
    $[3\rightsquigarrow\textsf{NETWORK}]$:1089 $\Rightarrow [5\rightsquigarrow\textsf{NETWORK}]$:900 & 0.83 \\
    \cellcolor{lightgray}$[8\rightsquigarrow\textsf{NETWORK}]$:1388 $\Rightarrow [5\rightsquigarrow\textsf{NETWORK}]$:1130 & 0.81 \\
    $[13\rightsquigarrow\textsf{LOG}]$:687 $\Rightarrow [13\rightsquigarrow\textsf{NC\_OUT\_STREAM}]$:556 & 0.81 \\
    \hline
  \end{tabular}
\vspace{-2ex}
\end{table}

Figure~\ref{fig:quantitative-risk} presents the app-level risk values of the apps in each cluster. For the convenience of the demonstration, the 9935 apps with no sensitive flow, whose risk value $R_x = 0$, are assigned with a minimal positive $10^{-4}$. We found the small clusters $c_2$ and $c_5$ are with high risk, because the major sink category \textsf{NC\_ICC} is risky and the average numbers of major sensitive flows (i.e., 210.85 per app in $c_2$ and 94.50 per app in $c_5$) in each app are big. Then followed by $c_6$, whose top-6 sensitive flows account for 55\% of all detected flows, and 41\% of them flow to the riskiest \textsf{NETWORK} sinks. The apps in cluster $c_7$ are generally at low risk because of the large portion of apps with no sensitive flows. In general, the ranges of risk values for different clusters discriminate well, which justifies the design of $R_x$ as well as the choices on severity weights in Table~\ref{tab:vul-weight} and risk weights in Table~\ref{tab:sink-weight}.

To figure out more evidence for the relations between important (\textit{vul\_type}, \textit{sink\_category}) pairs, we use the $\nu$ part of feature vectors to perform an association rule mining with the FP-growth method \cite{DBLP:conf/sigmod/HanPY00} provided by Weka \cite{DBLP:journals/sigkdd/HallFHPRW09}. The generated rules are presented in Table~\ref{tab:association}. We choose the associations with larger numbers of app instances ($> 500$) since we want to find associations between the top-ranking labels of each cluster. From the results, we found informative co-occurrences for the top-ranking labels of $c_1\sim c_4$. Moreover, the sensitive flows from insecure passwords for \textsf{KeyStore} (type 3) to \textsf{NETWORK} sinks are highly correlated with the sensitive flows caused by SSL/TLS-related vulnerabilities (type 4,5,8). The sensitive flows from SSL/TLS misuses, especially customizing \textsf{HostnameVerifier} and \textsf{TrustManager}, tend to happen simultaneously to trigger network leakages in many cases. Such correlations (type 3,8 to network) are also evidenced by the popular apps in Table~\ref{tab:real-world-findings}.

\vspace{1ex}
\noindent \framebox[\linewidth][l]{
\parbox{3.3in}{
Clustering-based assessment explains the most representative threats and estimates the risk level for each cluster, which can guide the vetting strategies of app stores. The frequently detected insecure PRNGs generally do not broaden the attack surface. The certificate-related and SSL/TLS-related misuses tend to trigger sensitive flows to the network simultaneously. Such aspects should be paid close attention to in developing apps.
}
}

\section{Discussion}\label{sec:discussion}

In this section, we discuss several further potential improvements and the threats to the validity of our approach.

\paragraph{Impact of the granularity of vulnerability types.} Constructing an adapter for a detector reporting some coarse-grained misuse type, e.g., constantError of \textsc{CogniCrypt}$_\textsc{sast}$, requires more effort. From the raw output of the detector, we extract the misuse tuples and the corresponding error type. We manually analyze over a small app set to build the mapping from each detected triple to some vulnerability $id$ of our comprehensive list. Although evaluated to be sufficient for our datasets, the small app set should be incomplete to deal with all the possible cases. Moreover, if we arrange more vulnerability types, e.g., use different vulnerability types to distinguish R3 and R6 of DiffCode, see column \textsf{DC} in Table~\ref{tab:rules}, we will have to do more manual analysis, even on the output of CryptoGuard. Therefore, the granularity of the vulnerability types matters to our approach.

\paragraph{False positives of analysis.}
We have introduced the voting mechanism of the detector chain to mitigate the impact of false positives of misuse detectors on real-world apps. However, the false positive of data-flow analysis is another issue. The accuracy of data-flow analysis impacts the assessment of the risk level of cryptographic misuses. In the current implementation, we faithfully adapt the results of FlowDroid into the risk assessment without addressing the inaccuracy of the data-flow analysis. Due to the homogeneous nature of state-of-the-art data-flow analyses on the taint-connection analysis, we may also bring in the voting mechanism based on multiple data-flow analyses, e.g., \cite{DBLP:conf/ccs/WeiROR14, DBLP:conf/pldi/ArztRFBBKTOM14}, to reduce the false positives of individual analysis.

\paragraph{Possible missing of flows.} When sensitive data threatened by cryptographic misuses first flows to an intermediate location with one sink and then read by another source to leak at another sink, such flow through the intermediate location should have been identified by the taint-connection analysis of \sysname{} because we enumerate both sinks for the taint-connection analysis. This issue becomes harder when the data-flow analysis is only modularly performed. Therefore \sysname{} requires a global data-flow analysis, and such analysis should be re-conducted when the app changes in evolution. Besides, there is no guarantee that the data-flow analysis can report all critical flows. The voting mechanism cannot solve such false negatives since voting is generally to decide the positives. Consequently, our approach is up against a potential underestimation of the cryptographic misuse risks.

\paragraph{Impact of API level and app popularity.} Android cryptographic APIs are subject to several changes in the higher API levels \cite{crypto-api-doc}, which affect the potential of specific cryptographic misuses on new Android systems. As a general-purpose cryptographic misuse assessment approach, \sysname{} focuses on any possible misuse but has yet to investigate the difference between the misuses at different API levels. However, such an investigation would be very helpful in guiding the development of the new cryptographic library APIs at new API levels; thus, we leave this for future work. Besides the severity weight $w_{id}$ and risk weight $w_{sc}$ used by \sysname{}, we can also adapt other weights, e.g., the popularity of the apps. Such a metric is not related to the app's attack surface but would be effective in a general-purpose risk assessment.

\section{Related Work}\label{sec:related}

Some works have addressed cryptographic misuse analysis through different means. The case study by Lazar et al. \cite{DBLP:conf/apsys/LazarCWZ14} analyzed the cryptographic vulnerabilities in CVE from 2011 to 2014. They found that over 80\% of cryptographic vulnerabilities were caused by misusing cryptographic libraries. In Android apps, the proportion of vulnerable apps containing cryptographic API misuse was 88\%, even under a straightforward threat model \cite{DBLP:conf/ccs/EgeleBFK13}. Other statistics show that over 65\% of iOS apps contain various defects caused by cryptographic misuses \cite{DBLP:conf/nss/LiZLG14}. Although people tend to simplify the design of cryptographic libraries by limiting the decision space of parameters, comprehensive documentation and code samples are critical \cite{DBLP:conf/sp/Acar0FGKMS17}, and misleading knowledge should be filtered out \cite{DBLP:conf/icse/MengNYZA18,DBLP:conf/icse/ChenFMWG19} on using the cryptographic APIs correctly. This intricate situation prompted us to investigate the vulnerability rules and the severity of cryptographic misuses in Android apps.

CryptoLint \cite{DBLP:conf/ccs/EgeleBFK13} is a pioneer work that proposes a static analysis to detect flows between improper cryptographic parameters and cryptographic operations. The authors proposed six common rules of cryptographic API usage against IND-CPA. BinSight \cite{DBLP:conf/ccs/MuslukhovBB18} considers the same set of rules, identifies the violations of these rules and tracks the sources of misuse with static program slicing. The results show that most misused call sites of cryptographic APIs originate from third-party libraries. CDRep \cite{DBLP:conf/ccs/MaLLD16} attempts to fix the cryptographic misuses identified with CryptoLint by instrumenting the app's bytecode with several patch templates. Their security rules slightly extend the rules of CryptoLint, i.e., \cite{DBLP:conf/ccs/EgeleBFK13, DBLP:conf/dasc/ShaoDGYS14}.

It has been realized that more fine-grained vulnerability specifications should correlate to the cryptographic API misuses \cite{DBLP:conf/pldi/PaletovTRV18, DBLP:conf/ecoop/KrugerS0BM18, DBLP:conf/ccs/RahamanXASTFKY19, DBLP:conf/msr/GaoKLBK19}. \textsc{CogniCrypt}$_\textsc{sast}$ \cite{DBLP:conf/ecoop/KrugerS0BM18, DBLP:conf/kbse/KrugerNRAMBGGWD17, cognicrypt, cognicrypt-android} takes the rules for correct usage of cryptographic APIs as input and translates them into a flow- and context-sensitive static analysis that checks apps for compliance with these rules. The rules are specified in the \textsc{CrySL} language. By defining the correct usages as a white list, \textsc{CogniCrypt}$_\textsc{sast}$ classifies the violations of these rules into seven misuse types \cite{cognicrypt}. CryptoGuard \cite{DBLP:conf/ccs/RahamanXASTFKY19} extends the threat model to include the misuse of SSL/TLS APIs. The authors developed an on-demand flow-, context- and field-sensitive program slicing and provided contextual refinements for false positive reduction. The data-flow analysis used in these works aims at identifying the misuse itself. In contrast, our misuse-originating data-flow analysis is to profiling the attack surface on the data threatened by the cryptographic misuses. On the temporal dimension, Paletov et al. \cite{DBLP:conf/pldi/PaletovTRV18} argued that cryptographic APIs and their exploits evolve over time. They derived the usage changes of cryptographic APIs from paired directed acyclic graphs based on mining code commit messages. A comprehensive set of security rules were elicited and evaluated over a dataset of Java projects. Gao et al. \cite{DBLP:conf/msr/GaoKLBK19} leveraged the report of \textsc{CogniCrypt}$_\textsc{sast}$ to find pairwise misuse updates among successive releases of the application. They found several evolutionary features indicating the inadequacy of the developer's fixing behaviors. The types and rules of cryptographic misuse in these approaches are more general than the threat model of CryptoLint. Compared with these cryptographic misuse detection approaches, our work aims at demonstrating the risk of dependency between cryptographic misuse and related data leakage. On the other hand, the general-purpose risk estimation approach \cite{DBLP:journals/dase/SonCF22} takes private data collection and sharing behaviors as the metrics to measure apps' risk. The data leakage modeling is relatively coarse-grained compared with the data-flow analysis used in this work. Our work focuses on the sensitive data operated by cryptographic APIs, which are more specialized than the personal data of \cite{DBLP:journals/dase/SonCF22}. Taking the information from Table~\ref{tab:rules}, we summarize the differences between related approaches in Table~\ref{tab:comparison}.

\begin{table}
\renewcommand{\arraystretch}{1.1}
  \caption{Comparison with Related Approaches}
  \label{tab:comparison}
  \begin{tabular}{l|c|c|c}
    \hline
    Approach & Vul. type & Data-flow & Risk \\
    & extent & analysis usage & assess \\
    \hline
    BinSight\cite{DBLP:conf/ccs/MuslukhovBB18} & low & N/A & N/A \\
    \textsc{CogniCrypt}$_\textsc{sast}$\cite{DBLP:conf/ecoop/KrugerS0BM18, DBLP:conf/kbse/KrugerNRAMBGGWD17} & medium & misuse identification & N/A \\
    CryptoGuard\cite{DBLP:conf/ccs/RahamanXASTFKY19} & medium & misuse identification & N/A \\
    Son et al.\cite{DBLP:journals/dase/SonCF22} & high & N/A & \checkmark \\
    This work & high & attack surface profile & \checkmark \\
  \hline
\end{tabular}
\vspace{-2ex}
\end{table}

Our work focuses on the impact of cryptographic misuses in general uses. In contrast, several works have addressed the impact of API-level cryptographic vulnerabilities on very specialized aspects, e.g., in ad libraries \cite{DBLP:conf/ccs/0001BD16}, and SSL/TLS \cite{DBLP:conf/ccs/FahlHMSBF12, DBLP:conf/ccs/GeorgievIJABS12, DBLP:conf/ndss/SounthirarajSGLK14, DBLP:conf/sp/FischerBXSA0F17}. MalloDroid \cite{DBLP:conf/ccs/FahlHMSBF12} performs static analysis on Android apps to detect different flaws related to the misuse of SSL/TLS, e.g., accepting all certificates or all hostnames. Georgiev et al. \cite{DBLP:conf/ccs/GeorgievIJABS12} analyzed the certificate validations in security-critical apps and libraries and found that the incorrect validations were caused by some critical design flaws of SSL/TLS APIs. SMV-Hunter \cite{DBLP:conf/ndss/SounthirarajSGLK14} focused on the custom validation code, i.e., overriding of \textsf{X509TrustManager} and \textsf{HostNameVerifier}. After detecting the custom validations and the UI entry points leading to these validations statically, its dynamic analysis emulates user interaction by feeding the app with customized inputs to trigger the vulnerable code under MitM attacks. Even in the vulnerability categories proposed by code smell and secure coding research, e.g., \cite{DBLP:conf/scam/GhafariGN17, DBLP:conf/ccs/NguyenWA0WF17, DBLP:conf/sp/FischerBXSA0F17}, cryptographic and SSL/TLS misuses are considered as a kind of major security vulnerabilities.

\section{Conclusion}\label{sec:conclusion}

To mitigate the struggle on common sense that all cryptographic misuses should be fixed mandatorily, we present \sysname{}, a framework combining adapter-based misuse detection and data-flow-driven risk assessment to estimate the threat of cryptographic misuses in Android apps. \sysname{} adapts different misuse detectors to achieve detections against more comprehensive vulnerability types, quantifies the risk of cryptographic misuses based on the result of the misuse-originating data-flow analysis, and predicts the potential threats of cryptographic misuses for the app vetting with unsupervised learning. We implemented an instance of \sysname{} and conducted evaluations on the accuracy of detection and the effect of risk assessment. With the help of \sysname{}, we observed security facts on the attack surface of apps and the severity of vulnerabilities caused by cryptographic misuses. We also found vulnerable popular real-world app whose data affected by cryptographic misuses tend to be leaked. Future work includes extending the sink categories and investigating the tendency of misusing different cryptographic libraries to benefit the development of cryptographic libraries.

CryptoEval has been made publicly available at \url{https://bitbucket.org/suncong_xdu/cryptoeval/src/master/}.

\bibliographystyle{IEEEtran}
\bibliography{IEEEabrv,mybib}

\end{document}